# Comparative Traffic Performance Analysis of Urban Transportation Network Structures

Behnam Amini[1], Farideddin Peiravian[2] (corresponding author), Morteza Mojarradi[3], and Sybil Derrible[4]

[1] Visiting Professor, University of Illinois at Chicago, Civil Engineering Department,
842 W. Taylor St., 2095 ERF (MC246), Chicago, IL 60607-7023
Tel: 312-996-3441, Fax: 312-996-2426, Email: behnama@uic.edu

[2] Research Assistant, University of Illinois at Chicago, Civil Engineering Department
842 W. Taylor St., 2095 ERF (MC246), Chicago, IL 60607-7023
Tel: 312-394-9020, Fax: 312-996-2426, Email: fpeira2@uic.edu

[3] Research Assistant, Imam Khomeini International University, Civil Engineering Department, Qazvin, Iran
Tel: +98-281-837-1122, Fax: +98-281-378-0073, Email: morteza20.mojaradi@yahoo.com

[4] Assistant Professor, University of Illinois at Chicago, Civil Engineering Department
842 W. Taylor St., 2095 ERF (MC246), Chicago, IL 60607-7023
Tel: 312-996-2429, Fax: 312-996-2426, Email: derrible@uic.edu

**Abstract**

The network structure of an urban transportation system has a significant impact on its traffic performance. This study uses network indicators along with several traffic performance measures including speed, trip length, travel time, and traffic volume, to compare a selection of seven transportation networks with a variety of structures and under different travel demand conditions. The selected network structures are: modified linear, branch, grid, 3-directional grid, 1-ring web, 2-ring web, and radial. For the analysis, a base origin-destination matrix is chosen, to which different growth factors are applied in order to simulate various travel demand conditions. Results show that overall the 2-ring web network offers the most efficient traffic performance, followed by the grid and the 1-ring networks. A policy application of this study is that the branch, 3-directional grid, and radial networks are mostly suited for small cities with uncongested traffic conditions. In contrast, the 2-ring web, grid, and 1-ring web networks are better choices for large urban areas since they offer more connectivity, thus allowing them to perform efficiently under congested traffic conditions.

# 1. Introduction

There have been many discussions within the realm of urban transportation planning and management about the impact of network structure and topology on travel patterns of people (Garrison and Marble, 1962; Kansky, 1963; Rodrigue, 2009). While some network structures are more suited for a given urban transportation system, others may not. Therefore, the question remains whether it would be possible to manage traffic in an urban transportation network through a proper choice of structural form for it, thus improving its traffic performance.

While there have been several studies on the relationship between network structure and travel demand parameters (Newell and Daganzo, 1986a, 1986b), only limited research has been done about the impact of network structure on urban traffic performance under varying travel demand conditions. An obvious reason for the lack of literature on this topic resides in the absence of case studies since different cities or neighborhoods with different network structures cannot be compared directly. This limitation warrants the need for studies based on analyses of reasonable hypothetical network structures and travel demands.

The overall objective of this research is to study the impacts of transportation network structures on five urban traffic performance measures: speed, trip length, travel time, traffic volume, volume to capacity ratio (v/c), and vehicle-distance travelled (VDT); and three network trip characteristics: maximum travel time, average travel time, and average trip length. It also involves a comparative analysis of these measures under different travel demand loadings, as well as comparison of some of their network indicators. In order to achieve the above objectives, the following steps will be taken:

i. Selecting and defining relevant traffic performance measures for the comparative analysis,

ii. Zoning of a hypothetical city and determining the zonal centers (centroids),

iii. Devising and measuring the properties of a variety of transportation network structures,

iv. Developing a hypothetical base origin-destination (O-D) matrix, as well as reasonable growth factors (to be applied to the base O-D matrix during the process),

v. Loading the O-D matrices and assigning the resulting traffic to the network,

vi. Performing comparative analysis of the selected traffic performance measures for different networks;

vii. Using network indicators to investigate the underlying characteristics of the selected networks that result in their differing traffic performance measures.

Thanks to this analysis, one will be able to propose appropriate network structure for a given urban system based on its characteristics. The rest of the paper goes as follows. First, the current literature on the subject is reviewed. Second, the data, materials and methods are introduced and detailed, where the seven networks are defined along with the OD matrix. Afterwards, the networks are analyzed under different demand conditions. Finally, the results are analyzed for each relevant traffic performance measure, followed by a more thorough discussion. Overall, we learn that while network structure possesses a significant impact, one type of structure consistently outperforms all others.

## 2. Literature review

A review of the literature on the subject of this study reveals a range of prior relevant research work, differing in their selected network structures, performance measures, their applications, and analysis methodologies (Polzin, 2004; Stead and Marshall, 2001). Several studies have focused on defining proper quantitative traffic performance measures for capturing the impacts of different transportation network structures (Buhl et al., 2006; Hu et al., 2008; Naga and Fan, 2007; National Research Council (U.S.) et al., 2010; Volchenkov, 2008; Xie and Levinson, 2007). Some other works have looked at the problem from a planning perspective, in which the interaction between the structure and spread of the network with travel parameters such as trip frequency and mode of travel have been studied (Jiang, 2007; Levinson, 2012; Rui et al., 2013; Yang et al., 2010; Zhang and Levinson, 2004). Several studies have also been performed on urban traffic network models and their correlation with the network structure (Crane, 1996; Geroliminis and Daganzo, 2008; Mazloumian et al., 2010; Tsekeris and Geroliminis, 2013). There have also been studies on network performance during major operational disruptions (Aderinlewo and Okine, 2009).

Crane (1996) developed a model for the relationship between urban form and travel demand, in which only travel time was considered. The study showed that differences in urban form parameters result in a variety of impacts on trip lengths and durations and that general trip cost or its variable cost are affected by travel modes.

Snellen *et al.* (2002)tried to model the relation between urban form, road network type, and mode choice for frequently-conducted activities. For that, they collected data for a variety of urban forms, road network types, and neighborhoods for some cities in Netherland, both at the disaggregated individual as well as the aggregated household levels. They developed models for three trip purposes, namely: work, shopping, and recreation. Those models included spatial and socio-economic parameters, as well as variables such as vehicle-distance travelled (VDT), total number of trips, trip lengths and number of trips for each vehicle type.

Xie and Levinson (2009) explored the topological evolution of transportation networks by using a simulation model firstly validated using empirical evidence and then applied to 16 idealized network structures such as ring, web, hub-and-spokes, and cul-de-sac. They also used measures of connectivity, density, heterogeneity, concentration, and connection patterns to study the temporal changes of topological attributes of those transportation networks. Their observation was that "typical connection patterns emerge in the networks; the spontaneous organization of network hierarchies, the temporal change of spacing between parallel links, and the rise-and-fall of places in terms of their relative importance are also observed, providing further evidence for the self-organization property of surface transportation networks" (Xie and Levinson, 2009). They also concluded that traffic/network measures could be used to quantitatively describe and compare complex transportation networks, thus showing significant benefits in urban and transportation planning.

Further research also exists that takes a complex network approach to address the question, notably on public transportation systems (Derrible and Kennedy, 2011, 2010, 2009; Derrible, 2012). Nonetheless, little work has investigated the impact of network structures on traffic-related performance parameters, hence the focus of this article.

## 3. Data, materials, and methods

### 3.1. Zoning

In this study, we defined a hypothetical city with 15 traffic zones, which we loaded with a synthesized base O-D matrix along with different growth factors. The city can be seen in Fig. 1.

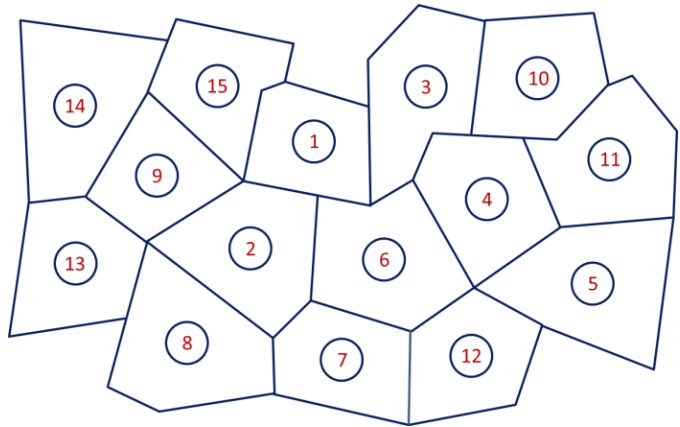

**FIG. 1 The hypothetical city and its zoning system.**

### 3.2. O-D Matrix

Table 1 shows the compiled base O-D matrix that was used. In order to examine the impact of different O-D loadings on the network, growth factors of 0.75, 1.25, 1.5, 1.75, 2.0, and 5.0 were selected and applied to the base O-D matrix. The resulting updated O-D matrices were then loaded onto the networks.

**TABLE 1 Base O-D Matrix**

| Zones | 1 | 2 | 3 | 4 | 5 | 6 | 7 | 8 | 9 | 10 | 11 | 12 | 13 | 14 | 15 |
|---|---|---|---|---|---|---|---|---|---|---|---|---|---|---|---|
| 1 | 0 | 4097 | 3650 | 5173 | 2374 | 3606 | 3189 | 2222 | 2974 | 2217 | 2264 | 1010 | 1658 | 1251 | 709 |
| 2 | 4144 | 0 | 2783 | 970 | 431 | 674 | 1352 | 2685 | 4236 | 430 | 379 | 308 | 1520 | 968 | 331 |
| 3 | 3652 | 2786 | 0 | 1056 | 218 | 233 | 568 | 506 | 1028 | 614 | 182 | 128 | 212 | 227 | 529 |
| 4 | 5120 | 959 | 1061 | 0 | 804 | 580 | 635 | 584 | 692 | 1607 | 2197 | 134 | 330 | 233 | 126 |
| 5 | 2345 | 444 | 216 | 862 | 0 | 1349 | 644 | 324 | 263 | 119 | 248 | 198 | 224 | 130 | 74 |
| 6 | 3539 | 668 | 229 | 566 | 1356 | 0 | 1200 | 274 | 506 | 146 | 272 | 499 | 215 | 164 | 41 |
| 7 | 3114 | 1403 | 542 | 611 | 606 | 1196 | 0 | 1007 | 1137 | 227 | 290 | 1265 | 542 | 466 | 140 |
| 8 | 2162 | 2758 | 515 | 575 | 298 | 278 | 1020 | 0 | 1598 | 225 | 200 | 244 | 869 | 444 | 90 |
| 9 | 2922 | 4205 | 1028 | 672 | 292 | 463 | 1085 | 1622 | 0 | 298 | 278 | 296 | 1664 | 1202 | 360 |
| 10 | 2178 | 432 | 597 | 1640 | 138 | 128 | 248 | 233 | 326 | 0 | 324 | 64 | 112 | 93 | 52 |
| 11 | 2226 | 365 | 198 | 2218 | 251 | 268 | 305 | 191 | 264 | 329 | 0 | 74 | 104 | 84 | 28 |
| 12 | 996 | 302 | 126 | 140 | 204 | 494 | 1250 | 236 | 290 | 50 | 86 | 0 | 180 | 162 | 32 |
| 13 | 1568 | 1538 | 222 | 305 | 248 | 181 | 534 | 890 | 1642 | 116 | 110 | 190 | 0 | 840 | 120 |
| 14 | 1238 | 1001 | 233 | 224 | 132 | 156 | 476 | 458 | 1211 | 97 | 92 | 170 | 833 | 0 | 293 |
| 15 | 692 | 346 | 520 | 126 | 70 | 41 | 134 | 89 | 370 | 49 | 28 | 28 | 120 | 299 | 0 |

## 3.3. Network Structures

The study includes seven different network structures: modified linear, branch, grid, 3-directional grid, 1-ring web, 2-ring web, and radial; as shown in Fig. 2. These structures were selected since they represent both real-life as well as theoretical networks. This aspect is particularly pertinent since many studies tend to remain either completely theoretical or case study based. Here, we tried to incorporate features from both realms.

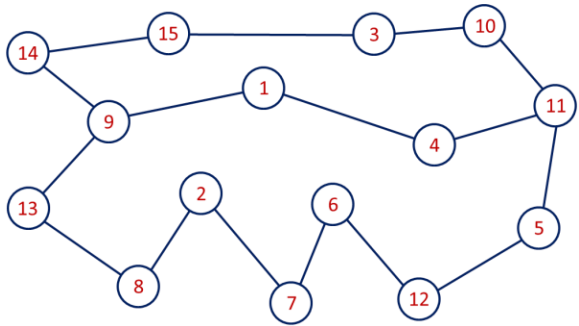

(a) Modified linear network (SM)

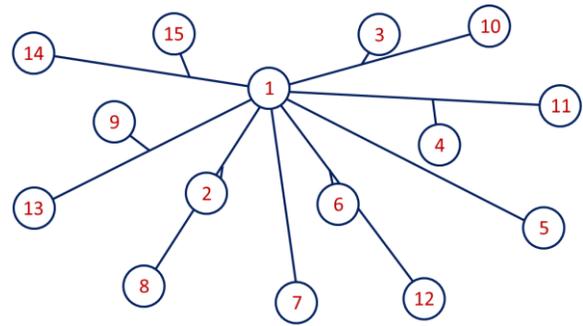

(b) Branch network (B)

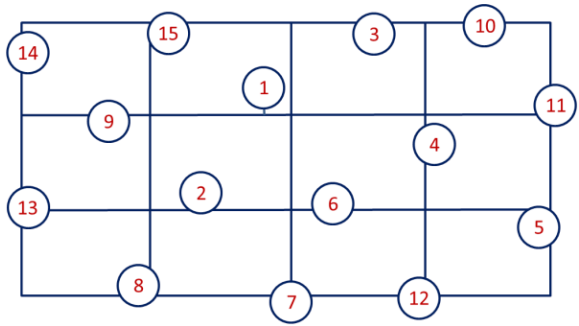

(c) Grid network (G)

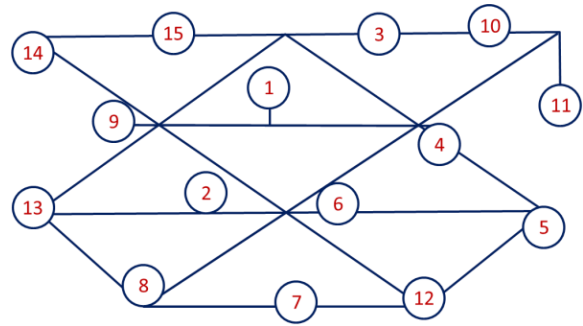

d) 3-Directional grid network (3DG)

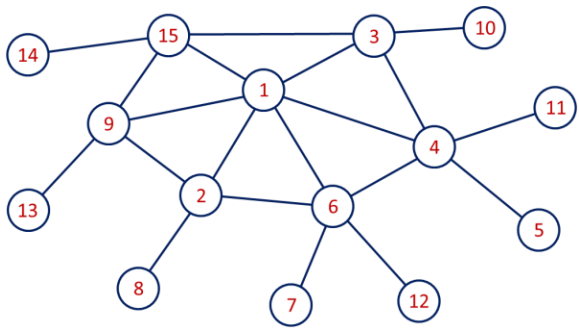

(e) 1-Ring web network (W)

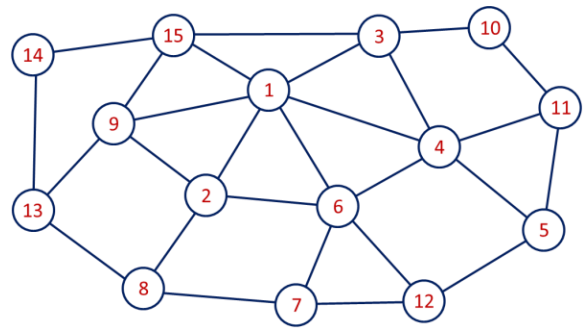

(f) 2-Ring web network (W2)

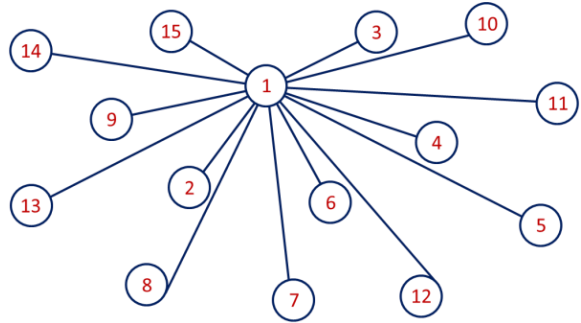

(g) Radial network (R)

**FIG. 2 Network structures.**

### 3.4. Network Indicators

Many indicators exist in the fields of graph theory and network science to characterize networks. Of the indicators initially used by Garrison and Marble (1962) and Kansky (1963), in this study we used three simple indicators that have been applied recurrently in the transportation realm: cyclomatic number $\mu$, 1$^{st}$ betti number $\beta$, and degree of connectivity $\gamma$.

For a network $G$ with $N$ nodes and $L$ links, the cyclomatic number ($\mu$) counts the number of cycles that are present in a network. It is mathematically defined as:

$$\mu = L - N + 1 \qquad \text{Eq. 1}$$

Essentially, $\mu$ counts the number of "extra" edges in the network as compared to a tree network. It is particularly relevant here since these cycles offer alternative routes, thus better distributing the traffic.

The betti number calculates the ratio of links to nodes. Mathematically, it is defined as:

$$\beta = L / N \qquad \text{Eq. 2}$$

The betti number ($\beta$) is also referred to as an indicator of complexity since networks with more links can be seen as being more "complex". It is lower than one for tree networks and higher than one for other networks. It can also be seen as the average number of links per node. A higher *β,* therefore, translates into more options to reach a destination.

Finally, the degree of connectivity (*γ*) is defined as the existing number of links divided by the total possible number of links in the network. The maximum total number of links in a network is ½$N(N-1)$, but because road networks are planar (i.e. two links crossing each other create a new node), the total possible number of links reduces to $3N - 6$. Mathematically, $\gamma$ is defined as:

$$\gamma = L / (3N - 6) \qquad \text{Eq. 3}$$

The Degree of connectivity is therefore bounded between 0 and 1, where completely connected networks take the value of 1. A higher degree of connectivity reflects the presence of more connections in a network, which also help better distribute traffic. For more detailed information about these three indicators, the reader is referred to Derrible and Kennedy (2011).

Table 2 shows the network indicator values for the seven networks studied here. As tree networks, the branch and radial networks have null cyclomatic numbers and betti numbers below one. On the other hand, the grid network has the most number of nodes as well as links. The 2-ring web network scores highest for all the three indicators calculated, which suggests it benefits from higher connectivity and complexity, while being more 'efficient' since it only has 28 links as opposed to 43 for the grid network. The grid network also benefits from a high cyclomatic number and relatively higher complexity and connectivity. The 1-ring web network trails behind the 2-ring web network for all three indicators.

Naturally, these different indicators represent network characteristics that affect traffic movement, thus traffic performance.

**TABLE 2 Network Characteristics of Seven Proposed Networks**

| Network | Nodes | Links | μ | β | γ |
|---|---|---|---|---|---|
| Modified Linear | 15 | 16 | 2 | 1.07 | 0.41 |
| Branch | 21 | 20 | 0 | 0.95 | 0.35 |
| Grid | 32 | 43 | 12 | 1.34 | 0.48 |
| 3-Directional Grid | 24 | 32 | 9 | 1.33 | 0.48 |
| 1-Ring Web | 15 | 20 | 6 | 1.33 | 0.51 |
| 2-Ring Web | 15 | 28 | 14 | 1.87 | 0.72 |
| Radial | 15 | 14 | 0 | 0.93 | 0.36 |

### 3.5. Traffic Assignment

During the traffic assignment step, the O-D matrices were loaded onto the selected transportation networks. The graphic software GNE 6.0 (Horowitz, 2013a) was used to draw the networks, where the nodes and links as well as their respective information were added. This information was then saved and used as input for the QRS II software (Horowitz, 2013b), which was used along with the moving average method. The maximum speed along all links was assumed to be 60 km/h and the link capacities were chosen to be 1200 veh/h.

The QRS II software (Horowitz, 2013b) uses the following formula in order to calculate the link travel times:

$$t = (1+\phi)t_b + \phi[t_0 + t_0\alpha(v/c)^\eta \sum_h p_h^{\eta+1}]$$  Eq. 4

where:

$\phi$ = step-size parameter;

$\alpha$ = volume/capacity multiplier;

$\eta$ = volume/capacity exponent;

$t$ = link travel time;

$t_b$ = base travel time;

$t_0$ = link free flow travel time;

$v$ = link volume;

$c$ = link capacity;

$p_h$ = probability that a vehicle trip assigned to the network occurs in the $h$-hour of the time period.

Eq. 4 can be further simplified into the following Eq. 5:

$$t = t_0 + t_0 \alpha (v/c)^\eta \sum_h p_h^{\eta+1} \qquad \text{Eq. 5}$$

where $\alpha$ and $\eta$ are constant coefficients defined in Eq. 4. Values of 0.73 and 2.1 were chosen for $\alpha$ and $\eta$, respectively, as a part of the study assumptions.

The outputs of QRS II software included: link traffic volumes (veh/h), link travel times (min), network average speed (km/h), maximum trip lengths (km), and the average trip lengths (km).

### 3.6. Performance Measures

In order to compare the performances of different networks, it was necessary to use relevant quantitative parameters. These measures were selected from the program outputs to properly represent the traffic conditions as well as network characteristics. As a result the following two groups of measures were selected:

i. descriptors of traffic characteristics: average speed, average volume, volume to capacity ratio (v/c), and vehicle-distance travelled (VDT);

ii. descriptors of network trip characteristics: maximum travel time, average travel time, and average trip length.

## 4. Results and analysis

The outputs of the QRS II program were used to calculate the selected traffic performance measures, while corresponding network indicator values were extracted from Table 2, all of which we use here to compare different network structures.

### 4.1. Analysis of average speeds

Fig. 3 exhibits the variation of average speed values for different network structures under various travel demand conditions.

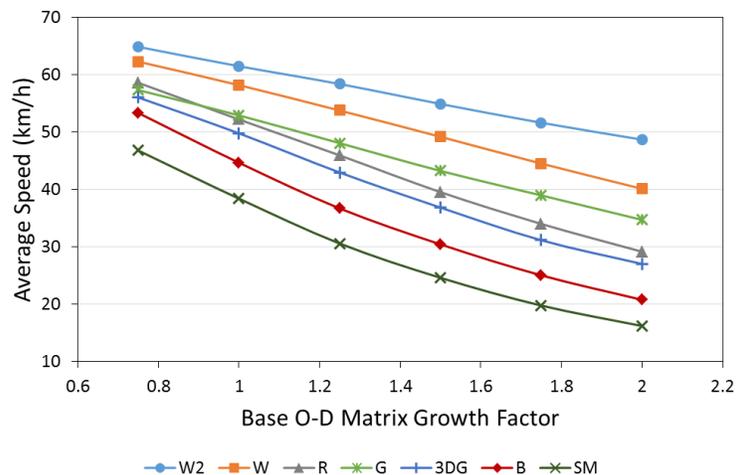

**FIG. 3 Variation of network average speed with respect to changes in travel demand.**

While the average speed shows mild nonlinearity trends for the branch and modified linear networks, for other networks it tends to have a negative linear relationship with travel demand. Moreover, from Fig. 3, the average speed for both 1-ring and 2-ring web networks are consistently higher than those of other networks. Indeed, akin to hub-and-spoke structures, web networks offer both radial and grid corridors. That manifests in their higher network properties, which provide easy access to the center and between the surrounding zones, hence resulting in lower overall congestion. Another interesting observation can be made between the purely radial and the purely grid network. While the radial structures exhibits higher average speeds for low volumes, this relationship is inversed for higher traffic volumes. This is particularly interesting since smaller urban areas, with low traffic volumes, tend to have grid networks, and large urban areas have radial expressway systems, which is in direct contradiction with our finding (not considering the web networks).

Fig. 3 also highlights the fact that on a larger scale there is a negative linear relationship between average speed and traffic volume. While this is expected, the capacity to absorb higher volumes has a significant impact on this trend. As a result, the trend is noticeably linear for the web networks, and some non-linear features start to appear for the branch and modified linear networks. Put simply, in the later cases, the traffic levels approach the critical density more quickly. This is further emphasized by the fact that the average speed values are consistently lower in these two networks as well.

### 4.2. Analysis of average traffic volumes

Fig. 4 presents the variations of average link traffic volumes in different networks with respect to changes in travel demand loadings. This indicator is calculated as the weighted average of traffic volumes among all network links.

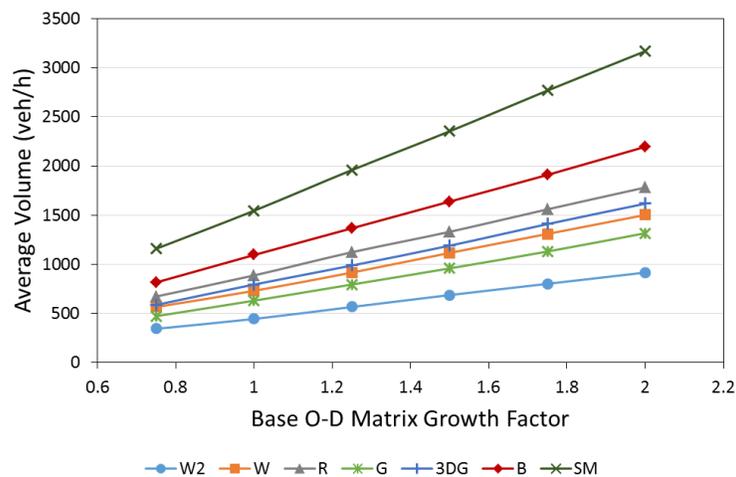

**FIG. 4 Variation of network traffic volume with respect to changes in travel demand.**

As opposed to the previous section, the main observation from this Fig. 4 is the linear trend regardless of network structure. Because route choice options are limited in the modified linear, radial, and branch networks, it is understood that incremental increases in the traffic volumes will continue to be assigned to the previously selected routes. As a result the traffic volume changes for these networks are expected to be linear with respect to the demand changes. The performance of the web networks is also interesting. Indeed, despite having significantly fewer links compared to the grid network, they perform as well or even better for the 2-ring web network. Overall, Fig. 4 demonstrates that networks that distribute traffic more efficiently, such as 2-ring web network, also experience lower average traffic volumes.

**4.3. Analysis of volume-to-capacity ratios (*v/c*)**

The volume-to-capacity ratio, *v/c*, is used extensively in traffic engineering, in particular to classify the road level of service. Fig. 5 presents the variations in the *v/c* ratios as a function of travel demand for the various network configurations.

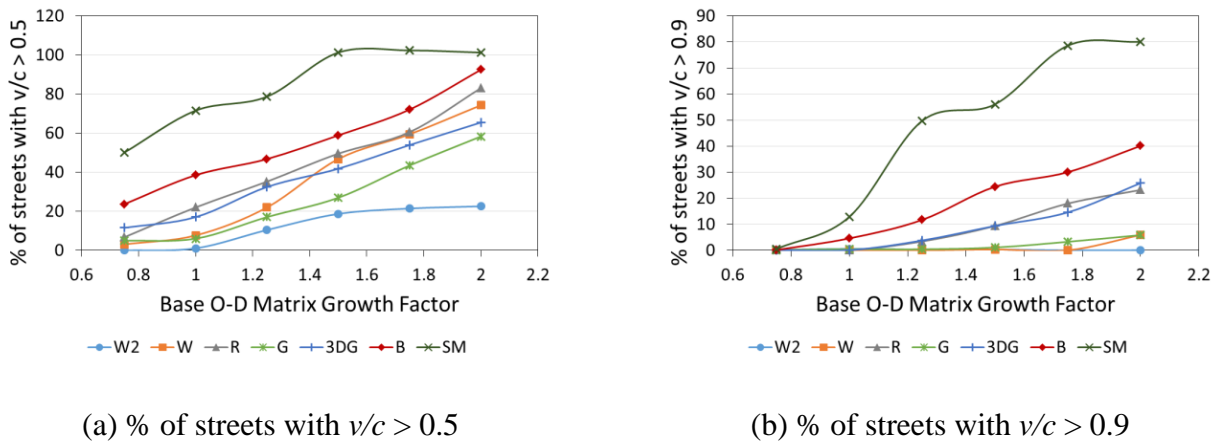

(a) % of streets with *v/c* > 0.5          (b) % of streets with *v/c* > 0.9

**FIG. 5 Variation of % of streets with (a) *v/c* > 0.5 and**

**(b) *v/c* > 0.9 with respect to changes in travel demand.**

From Fig. 5(a), the percentage of streets with *v/c* > 0.5 increases under higher demand conditions, which is expected. Nevertheless, for the modified linear and radial networks the graphs behave irregularly as the loaded travel demand increases. The reason is that since those networks offer fewer opportunities for the traffic to distribute, an increase in the demand will first affect the currently used links more than other links. As the demand increases, the rates of changes of increase in link volumes do not stay constant and will vary irregularly as the traffic spills over to other links. In contrast, the plots for all other networks, except the 2-ring web, show regular and relatively consistent trends, as any additional demand is distributed over the network efficiently. By far, the 2-ring web network performs best since its volume-to-capacity ratio never goes above 20%, i.e. the percentage of semi-congested streets stays low in that network.

Fig. 5(b) shows the percentage of streets with $v/c > 0.9$. The value 0.9 was selected because it represents close-to-critical conditions when the links are under significant stress and the levels of service are very low, i.e. the networks will be performing inefficiently. From Fig. 5(b), the $v/c$ ratios for all networks increase as the demand increases, while staying lower than those of $v/c > 0.5$, as expected. The changes in the $v/c$ ratios with respect to changes in demand in Fig. 5(b) also show similar trends to those of Fig. 5(a), except for the 2-ring web, 1-ring web, and grid networks, for which the traffic distributions is more efficient as there are hardly any streets with $v/c > 0.9$. In fact, these results demonstrate that, thanks to their network structure, the 2-ring web, 1-ring web, and grid networks are the most efficient in terms of level of service. Other networks, especially the modified linear network, do not perform as well.

### 4.4. Analysis of vehicle-distance traveled (VDT)

Fig. 6 shows the variations of VDT (in veh-km) for the selected network structures with respect to changes in travel demand loadings.

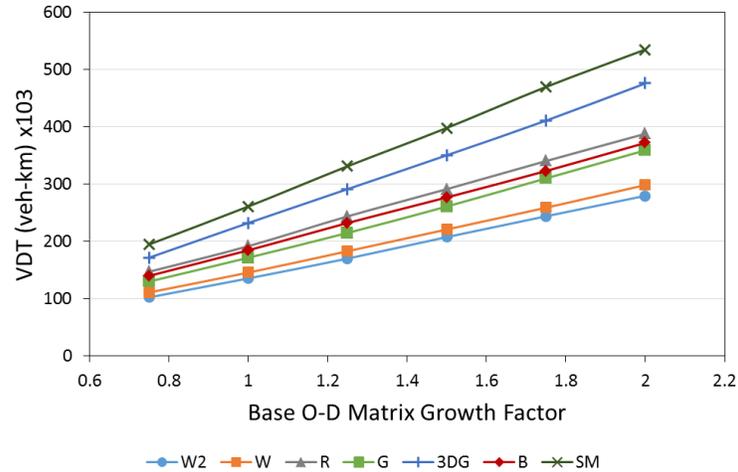

**FIG. 6 Variation of VDT with respect to changes in travel demand.**

In Fig. 6, VDT values for all networks vary linearly with an increase in travel demand. Indeed, while an increase in demand will negatively and non-linearly impact travel times, it does not affect the traveled distances (i.e. link lengths). In fact, the behavior observed in Fig. 6 is similar to Fig. 4, something that is expected since VDT is simply the product of link volumes and link lengths. Overall, the 1-ring and 2-ring web networks show the lowest VDT values, the grid, branch, and radial networks fall in the middle, and finally the 3-directional grid and modified linear networks show the highest VDT values.

## 4.5. Analysis of maximum and average travel times

The variations of maximum travel times (in min) for the selected network structures with respect to changes in travel demand loadings are shown in Fig. 7 (a).

Fig. 7(a) shows that the maximum travel time values vary non-linearly with changes in travel demand, something that is expected due to the non-linearity observed in Eq. 1 and Eq. 2. Similar to the earlier observations, the 2-ring, 1-ring and grid networks outperform the other networks, again thanks to their network characteristics that facilitate better, thus more efficient, distribution of the additional traffic.

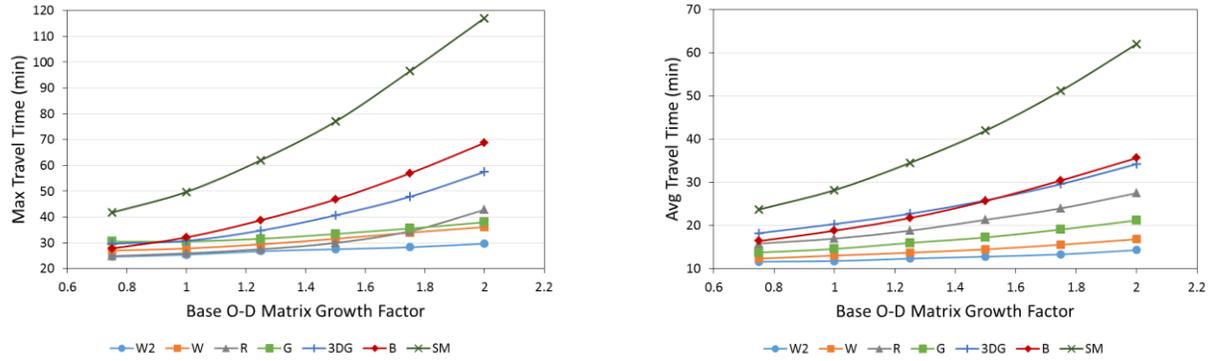

**FIG. 7 Variations of (a) maximum and (b) average travel times with respect to changes in travel demand.**

From Fig. 7(a), the radial network shows an atypical behavior. While it is first performing as well as the 2-ring web network, its travel time increases suddenly for demand growth factor values larger than 1.5. The reason for such behavior is that while the radial network can handle low levels of traffic volume rather efficiently, it is not capable of doing the same for higher loaded demand values due to the fact that its structure cannot offer the opportunity for the traffic to be diverted from the high-volume links to the low-volume ones, a weakness directly related to its tree structure (manifested by its null cyclomatic number). The remaining networks exhibit increasing operational inefficiency as the demand increases, resulting in stronger nonlinear trends.

Fig. 7(b) shows the variations of average travel times (in min) for the selected network structures with respect to changes in travel demand. As expected, and similar to Fig. 7(a), Fig. 7(b) shows increasing trend in average travel times as a result of increase in travel demand. Similarly, the 2-ring and 1-ring web networks, followed by the grid network, outperform the other networks, markedly thanks to their ability to better distribute additional increases in demand over their links. The radial network, however, does not exhibit the unexpected trend it followed in Fig. 7(a). Indeed, taking the average of travel times (as opposed to maximum travel time) among the links in a given network incurs a smoother behavior since sudden changes are toned down. For other networks, the rates of increase in average travel time are stronger as the demand increases, especially for growth factors more than 1.25.

### 4.6. Analysis of average trip length

The variations of average trip lengths (in km) for the selected network structures with respect to changes in travel demand loadings are shown in Fig. 8.

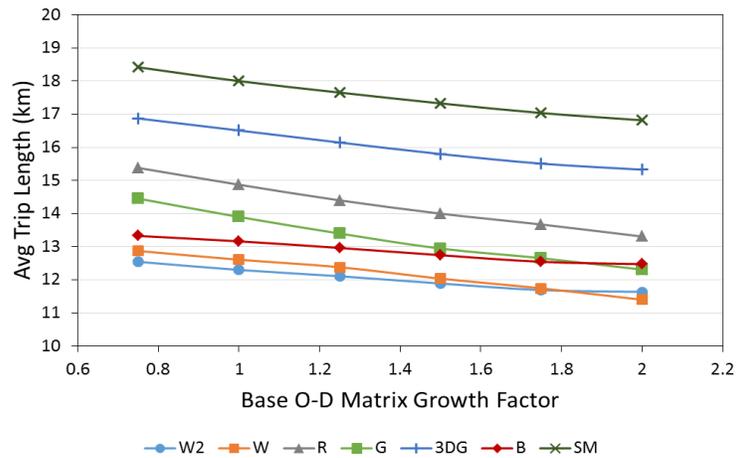

**FIG. 8 Variation of average trip length with respect to changes in travel demand.**

Fig. 8 shows that for all the selected networks, the average trip length decreases as the demand increases. This is due to the fact that as the traffic volume increases, the total distance traveled increases as well but at a slower pace than the demand or number of trips. As a result, the overall average trip length (i.e. the total distance traveled divided by the total number of trips) decreases. Nonetheless, here again, the 2-ring and 1-ring web networks perform better than the other networks; it is also worth noting that the 1-ring web network outperforms the 2-ring web network under critical conditions simply because the peripheral second ring incurs longer distances. Moreover, interestingly, the branch network shows a better performance than the grid network with respect to the average trip length values.

## 4.7. Analysis of critical conditions

In the previous sections, common levels of travel demand were simulated to analyze the performances of different networks structures. In order to study extreme situations, where the traffic conditions in the networks reach critical levels (i.e. system break-down), an additional scenario was simulated in which the base O-D matrix was multiplied by a growth factor of 5. Fig. 9(a) shows the variations of average travel times for the networks as the travel demand increases for both the previous O-D matrix growth factors as well as the new one.

As expected, average travel time consistently increases with travel demand. This is clearly shown in Fig. 9(a), where all networks show increasing upward trends as the traffic volumes increase. In concomitance with our previous findings, the 2-ring web and 1-ring web networks perform the best, followed by the grid network. Similarly, the modified linear network, followed by the branch network, performs worse.

Additionally, the percentages of streets with $v/c > 0.9$ were also plotted in Fig. 9(b).

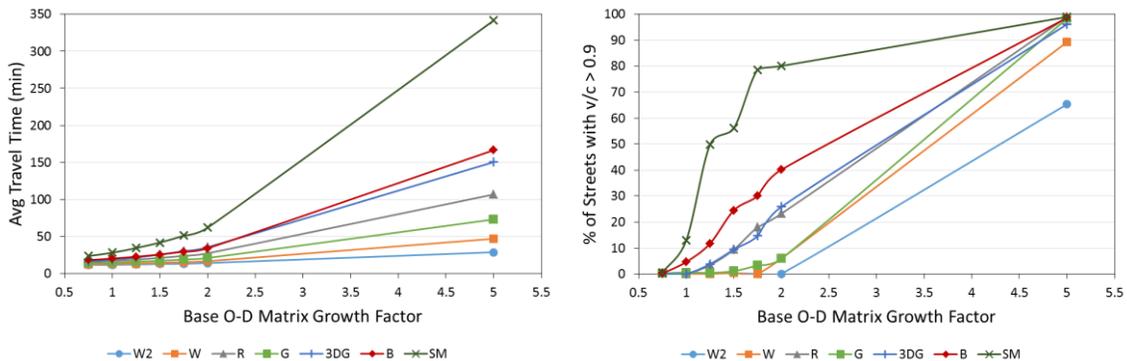

**FIG. 9 (a) Variation of average travel time versus changes in travel demand,**

**(b) Variation of % of streets with $v/c > 0.9$ versus changes in travel demand.**

Interestingly, while the 2-ring web network continues to perform the best at all levels of travel demand, the 1-ring web network fails to continue outperforming the grid network at high demand levels. In fact, this demonstrates that the grid network can handle and distribute additional demand more efficiently than the 1-ring web network as the traffic conditions approach critical levels. What is more, the 1-ring network actually tends to perform as poorly as other networks (except the 2-ring and grid networks) at critical demand levels.

## 5. Discussion

Having the above analyses results in mind, relating the results to the specific network indicators can provide more insights into the performance of those networks.

From Table 2, the grid network has the most number of links and nodes, followed by the 3-directional grid and 2-ring web networks. In contrast, the modified linear and radial networks have the least number of links and nodes. As a result, since the grid network provides more alternatives for traffic movement, it is expected to perform the best among the selected networks. This, however, is not completely accurate as was observed and discussed in the comparison of traffic performance measures. The phenomenon can be best understood by looking at the three network indicators introduced above (cyclomatic number, betti number, and degree of connectivity).

Regarding the cyclomatic number, the 2-ring web and grid networks have the highest values, while the branch and radial networks have null cyclomatic numbers $μ$. Again, the cyclomatic number counts the number of "extra" edges compared to a tree network, which essentially create cycles and build redundancies. Since these cycles offer alternative routes, networks with higher cyclomatic numbers are expected to better distribute the traffic, something which is also confirmed by previous observations.

Regarding the 1$^{st}$ betti number $β$, the 2-ring web has the highest $β$, while the branch and radial networks have the lowest values. As the average number of connections per node, a higher *β* essentially translates into more options to reach a destination, which is a desirable feature. This explains the weak performance of the branch and radial networks.

Finally, regarding the degree of connectivity *γ*, the 2-ring web network has the highest value, followed by the three closely ranked 1-ring web, 3-directional grid, and grid networks. In contrast, the branch and radial networks perform the worst with respect to this indicator. Such an observation strengthens the previous findings via traffic performance measures.

Overall, it is clear from the results that the 2-ring web network performs best among all networks. That is due to the fact that by its very nature, the 2-ring web network contains relevant features from both radial and grid networks. Indeed, its radial elements allow traffic to be distributed efficiently, minimizing distance traveled, while its grid elements allow for alternatives to better distribute the traffic under congested conditions. This duality seems to be the critical feature that helps it outperform the other networks.

## 6. Conclusion

This article focused on the performance analysis of a variety of network structures in response to increase in travel demand. For that, several traffic performance measures were selected and calculated for each network under increasing travel demand. Some relevant network indicators were also calculated for the same networks. The results were then compared and analyzed.

The analysis showed that under uncongested conditions, all networks performing similarly, with the possible exception of the modified linear network. As traffic volume (i.e. congestion) increased, however, and the conditions approached critical levels, the various networks started to show significant differences in their performance.

Overall, the seven networks analyzed in this study can be categorized into the following three groups with respect to their performance under varying demand conditions:

Group 1: modified linear network;

Group 2: branch, 3-directional grid, and radial networks;

Group 3: 2-ring web, 1-ring web, and grid networks.

Group 1 network consistently showed the worst performance over the range of all measures used for comparative analysis.

Group 2 networks showed varying behaviors. While under uncongested conditions the 3-directional grid performs better, it is the radial network that outperforms the other two under congested conditions.

Among Group 3 networks, and in fact among all the selected networks, the 2-ring web networks performed the best under both uncongested as well as congested conditions, with respect to all traffic performance measures. The other two networks, i.e. the 1-ring web and grid networks, outperformed each other alternatively under various conditions and with respect to different performance measures, while still performing better than the networks in the other two groups.

A possible policy application of this study is that Group 2 networks may be better suited for small cities if the traffic conditions stay mostly uncongested. In contrast, it is Group 3 networks that are best suited for large cities since their structures better absorb and distribute additional demand via their links.